\pdfoutput=1
\documentclass[12pt]{article}
\usepackage{amsmath}
\usepackage{graphicx}
\usepackage{float}
\usepackage{enumerate}
\usepackage{natbib}
\usepackage{cleveref}
\usepackage{url} 

\newcommand{\blind}{0}

\newcommand{\ve}[1]{\mathbf{#1}}
\DeclareMathOperator*{\argmin}{argmin}

\begin{document}

\defcitealias{bcs}{\textit{Bowl Championship Series} n.d.}
\defcitealias{usau_club_guidelines}{\textit{USAU Club Guidelines} 2020}
\defcitealias{usau_2021}{\textit{Club Division} 2021}
\defcitealias{usau_rankings}{\textit{Team Rankings} 2018}

\def\spacingset#1{\renewcommand{\baselinestretch}%
{#1}\small\normalsize} \spacingset{1}


\if0\blind
{
  \title{\bf An Empirical Study of Least Squares Ratings for USA Ultimate Frisbee}
  \author{Alexander N. Sietsema\hspace{.2cm}\\
    Department of Mathematics, Michigan State University, East Lansing}
  \maketitle
} \fi

\if1\blind
{
  \bigskip
  \bigskip
  \bigskip
  \begin{center}
    {\LARGE \bf An Empirical Study of Least Squares Ratings for USA Ultimate Frisbee}
\end{center}
  \medskip
} \fi

\bigskip
\begin{abstract}
    Ultimate frisbee is one of the fastest-growing sports in the world. In the United States, the governing body USA Ultimate uses a custom power rating system to determine bid allocations for various competitive tournaments. However, this rating system has significant flaws and leaves room for improvement. In this paper, we apply the least squares rating system and demonstrate its improvement over the current system both qualitatively and relative to a number of quantitative metrics.
\end{abstract}

\noindent%

\noindent {\it Word Count:} 4018
\vfill

\newpage
\spacingset{1.45} 

\section{Introduction}

Ultimate Frisbee (hereafter ``ultimate\footnotemark'') is one of the fastest-growing sports in the world.
Ultimate has over three million participants in the United States alone, and is played in more than 100 countries worldwide \citep{ultimate_participation}.
In 2015, the World Flying Disc Federation was granted full recognition by the International Olympic Committee \citep{wfdf_2020}. 
Ultimate is played at the youth, college, and club levels.

\footnotetext{As the term ``Frisbee'' is a registered trademark, the sport is referred to as simply ``Ultimate'' in competitive settings. We will adopt this terminology.}

Ultimate is a non-contact disc sport played by two teams of seven players. 
The objective is to score goals by advancing the disc into the opposing team's end zone.
Players are not allowed to move while holding the disc, so the disc must be advanced by being thrown between players.
If the disc is dropped or is intercepted by the opposing team prior to entering the end zone, the opposing team takes possession and attempts to score.
Possessions continue to alternate until one team scores a goal. 
Ultimate combines elements of football, soccer, and basketball into a dynamic, fast-paced sport.
Ultimate is generally divided into three gender divisions: men's/open, mixed, and women's, each depending on the gender identification of the players involved.

Ultimate in the United States is governed by USA Ultimate (USAU), and is responsible for the national organization of youth, college, and club tournaments and championships. 
At the club level competition is structured through the Triple Crown Tour, in which over 600 teams participate in sectional and regional tournaments to win a bid to the National Championships \citepalias{usau_2021}.
In order to determine some of the bids to these tournaments, USAU employs a custom power rating system to gauge the relative strength of each team in competition \citepalias{usau_club_guidelines}. As in other sports, these ratings are the source of much discussion even outside of their tournament implications.

The problem of ranking sports teams is nearly as old as sports themselves. 
The question ``Which team is best?'' is a common topic of discussion for fans and analysts alike, and its answer can have a lasting impact on the landscape of the sport, the teams involved, and the fans who cheer them on.
Accordingly, many kinds of mathematical ranking systems have been created for nearly every sport.
A representative overview of such methods is beyond the scope of this work---for this reason, and for reasons we will describe below, we will focus mainly on rankings for football, particularly for the National Collegiate Athletic Association Football Bowl Subdivision (NCAAF FBS). \citet{stefani1997survey} provides a relatively extensive overview of systems across all sports.

Rankings for FBS college football tend to be controversial. Because teams play relatively few games and there is not an expansive postseason in the FBS, it is difficult to draw well-informed conclusions from the available data. Further, different teams play schedules of wildly different strengths, which introduces additional uncertainty about comparisons between teams.

A variety of mathematical approaches have been taken to solve this problem. The Colley Matrix method \citep{colley2002colley} ignores margin of victory and attempts to construct an unbiased set of rankings based on solely win/loss record and strength of schedule. This method is one of the most popular, and was used as one of the computer selectors in the Bowl Championship Series (BCS) \citepalias{bcs} system from 1998 to 2014. 
\citet{massey_ratings} created perhaps the other most popular computer system for college football, and was also used as a BCS selector. 
Massey's method takes final score, date, and location (home or away) into account, but uses a significantly more complicated Bayesian approach to produce a set of appropriate rankings.  \citet{chartier2011sensitivity} compared these methods to the Markov/PageRank method \citep{page1999pagerank} and concluded that both the Colley and Massey methods demonstrate better stability properties than the Markov method in some settings.

Other commonly discussed approaches include Sagarin's rankings \citep{sagarin_ratings}, the Billingsley Report \citep{billingsley_rankings}, the Football Power Index \citep{inside_fpi}, and Bill Connelly's SP+ \citep{explaining_sp}.
Kenneth Massey keeps a comprehensive list of over 100 different ranking methods for FBS football on his website \citep{massey_composite}.

Of most importance to us is the least squares method.
The least squares method for sports ratings was first introduced for football and basketball by \citet{stefani1977football, stefani1980improved}, but was popularized by Kenneth \citet{massey1997statistical}. 
As we will describe below, the least squares method attempts to produce ratings that respect the point differential of each game as closely as possible by solving an equation which minimizes the least squares distance between predicted and observed game outcomes.

The efficacy of the least squares method in sports has been studied extensively.
\citet{gill2009assessing} performed a survey of rating methods for college football including ordinary and weighted least squares.
\citet{barrow2013ranking} compared the predictive capabilities over six sports of eight ranking methods, including least squares, and showed that least squares performed significantly better than other methods on college football data.

The best choice of rating method for a given sport is dependent on the gameplay and structural properties of the sport.
Structurally, the Triple Crown Tour for ultimate is similar to FBS college football in that teams tend to play relatively few games in the regular season.
However, the most important factor distinguishing ultimate from other sports is the conditions that end a game. Rather than using a time limit, ultimate is usually played to a fixed goal cap, often 15 goals.
This makes score differential the most important metric in determining the relative strength of two teams in a given game, compared to a sport like football where a low-scoring game may have a different interpretation than a high-scoring game.
Additionally, this means that it is impossible to run up the score to the same degree as in other sports. This consideration with the least squares method was noted in \citet{gill2009assessing}.

This goal cap is not always 15 points, however, and may change depending on a number of factors. 
The goal cap may simply be chosen according to the tournament, or may be dependent on extenuating circumstances in a given game (e.g., inclement weather).
In order to limit the potential length of a game, soft and hard time caps are employed. 
Once the soft time cap is reached, the current point is finished, and the goal cap is set at one more than the current winning score.
If the hard time cap is reached, the current point is finished, and the team with the higher score wins.
If the score is tied, a final sudden-death point is played to determine the winner.
These complications may affect the final score of a game.

Currently, USAU uses an iterative power rating method designed to take into account some of these factors.
Score differentials are used as the basis for comparisons, though they are significantly altered through a formula. 
Ratings are constructed using an iterative weighted averaging process taking into account the goal cap and date. 

Aside from USAU, very little work has been done on ultimate ratings. Cody Mills maintains a website \citep{frisbee_rankings} which lists USAU and probabilistic bid ratings for the college and club divisions. \citet{goexploring} also discusses the problem of predicting the success of ultimate teams, but focuses mainly on player data rather than rankings.

This paper is organized as follows: in Section \ref{sec:methods}, we will further discuss the current power rating method used by USAU as well as the least squares estimator for sports ratings.
Section \ref{sec:empiricalcomparisons} will provide direct performance comparisons of each rating method on recent data from the Triple Crown Tour. 
Finally, in Section \ref{sec:conclusion}, we will provide closing remarks as well as ways in which the least squares method may be improved in the context of ultimate.

\section{Methods} \label{sec:methods}
\subsection{The current method}\label{sec:currentmethod}

We summarize from \citepalias{usau_rankings} the rating system currently used by USAU.
For each game played, a team earns a \textit{game rating} calculated based on the score differential of the game and adjusted based on the team rating of the opponent. 
These game ratings are calculated for each eligible game on a given team's schedule, and are aggregated into a \textit{team rating} through a weighted averaging process.
Each team begins with a team rating of 1000, and this process is iterated to convergence to produce the set of final team ratings.

The game rating is calculated as follows:

\begin{equation}\label{eq:usau}
   G_r = T_r \pm \left(125 + \frac{475}{\sin 0.4\pi} \cdot \left(\sin \left(\min\left(1,2 \cdot \left(1-\frac{l}{w - 1}\right)\right)\right) \cdot 0.4\pi\right) \right),  
\end{equation}
where $T_r$ is the team rating of the opponent, $G_r$ is the resultant game rating, and $w$ and $l$ are the winning and losing scores, respectively.
The sign of the second term is positive for a win and negative for a loss.

The explanation for this choice of function is given in \citepalias{usau_rankings}:

\begin{quote}
    The function was chosen to have the following properties:
    \begin{itemize}
\item Each additional goal is worth more when games are close than when they are not.
\item Every game decided by one point gets the same differential of 125, no matter the game total.
\item The maximum x [the right term of \eqref{eq:usau}] can be is 600.
\item A game earns the maximum differential if and only if the winning score is more than twice the losing score.
\end{itemize}

\end{quote}

Weighted averaging is then applied to the set of game ratings for each team to determine a final rating.
The weights are dependent on the date the game was played as well as the final score.
The date weight is given by the formula 

\begin{equation}
    d_w = 2^{\left(\frac{t}{n}\right)-1}
\end{equation} where the game is played in the $t^{th}$ week of an $n$-week regular season.
This function is chosen to give smaller weights to games earlier in the season. 
The score weight is given by the formula
\begin{equation}\label{eq:scoreweight}
    s_w = \min\left(1, \sqrt{\frac{w + \max(l, \lfloor \frac{w-1}{2} \rfloor)}{19}}\right),
\end{equation}
where $w$ and $l$ are again the winning and losing scores, respectively. 
This function is chosen to assign smaller weights to games played to fewer than $13$ points and with a combined score of both teams fewer than $19$ points.
Effectively, this score weighting assigns smaller weights to games with unusually small goal caps.
The final weighting for a game is the product of the date and score weights.

As above, the final set of power ratings is produced iteratively: each team is given an initial rating of $1000$, and this game rating calculation and weighted averaging process is iterated to convergence.

There are also conditions set on which games are eligible to be counted even among sanctioned regular-season tournaments.
Teams are required to play at least 10 games in order to be ranked, and games played by teams with ineligible rosters are not considered in the rating calculations. 
One more condition is set depending on score differential and the strength of the teams involved \citepalias{usau_rankings}:

\begin{quote}
Finally, if a team is rated more than 600 points higher than its opponent, and wins with a score that is more than twice the losing score plus one, the game is ignored for ratings purposes. 
However, this is only done if the winning team has at least $N$ other results that are not being ignored, where $N=5$. 
\ldots [this] removes the possibility that a team rated more than 600 points higher than its opponent will drop in rating when beating that team by a large enough point differential.
\end{quote}

Overall, this rating system is reasonably effective at producing rankings that agree with an ``eye test'' comparison of teams, but it also has significant drawbacks.
The formulas chosen for the game ratings and the weights are totally arbitrary up to the few listed conditions for each.
It is not clear, for example, why in the game rating calculation a sine function is chosen over another function, nor why the exact behavior of the square root function chosen for the score weight is correct. 

Additionally, the game eligibility restrictions based on score differential are unusual, and, as we will demonstrate below, may not be necessary in order to construct effective ratings.

\subsection{Least squares ratings}\label{sec:newapproach}

While relying on similar underlying ideas, the least squares method is significantly simpler than the USAU method. 
The least squares method creates a set of ratings that try to capture the expected point differential between two teams. 
In particular, we assume that the final score differential of a game is equal to the linear difference in rating between the teams involved. Each game of the season then represents an equation relating the rating of two teams by their score differential; treating these equations as a system and solving yields a set of ratings for each team.

For a brief example, consider a three game season:
\begin{table}[H]
\centering
\begin{tabular}{c}
Team A defeats Team B $15-10$\\
Team A defeats Team C $15-2$\\
Team B defeats Team C $15-7$\\
\end{tabular}
\end{table}
In this case, taking the score differentials as the differences in rating, we can create the linear system
\begin{align*}
    r_A - r_B &= 5\\
    r_A - r_C &= 13\\
    r_B - r_C &= 8.
\end{align*}

Solving this system yields the ratings $r_A = 6, r_B = 1, r_C = -7$. 
These ratings reflect our intuition about these teams: Team A won both games decisively and is certainly the best, Team B lost to Team A but beat Team C and is in the middle, and Team C lost badly twice and is the worst.
However, this solution will only be unique up to some constant shift, as adding a constant to all ratings still preserves their differences. 
To account for this, we add an additional equation which specifies the sum of the ratings of all teams to be zero.

In the general case, we may view this system as the matrix equation
\begin{equation}
    \begin{bmatrix}
    1 & -1 & 0 & \cdots\\
    1 & 0 & -1 & \cdots\\
    0 & 1 & -1 & \cdots\\
    \vdots & \vdots & \vdots & \ddots
    \end{bmatrix}
    \begin{bmatrix}
    r_A \\ r_B \\ r_C \\ \vdots
    \end{bmatrix}
    =
    \begin{bmatrix}
    b_1 \\ b_2 \\ b_3 \\ \vdots
    \end{bmatrix}
\end{equation}
where the schedule matrix $A$ has dimensions $m \times n$, where $m$ is the number of games played in the season and $n$ is the number of teams in the league, the rating vector $\ve{r}$ has values that represent the ratings of each team, and the score differential vector $\ve{b}$ has values that represent the score differential of a given game. 
A given row (game) will have a value of $1$ in the column corresponding to the winning team and a $-1$ corresponding to the losing team, with zeros elsewhere for the uninvolved teams. 
The entry of $\ve{b}$ in the same row will have the positive score differential of the game.
The shown values correspond to the example given above with $b_1=5$,  $b_2=13$, $b_3=8$.

To produce a set of ratings, we need only solve this matrix equation. 
Of course, in practice there is almost never a true solution to this equation.
For the least squares method, we use the ordinary least squares estimator

\begin{equation}
\hat{\ve{r}} = (A^\top A)^{-1} A^\top \ve{b}
\end{equation}
which solves the minimization problem
\begin{equation}
    \hat{\ve{r}} = \argmin_{\ve{r}} \|A\ve{r} - \ve{b}\|_2^2
\end{equation}
To take into account the difference in score caps, we apply a preprocessing step which multiplicatively normalizes all of the score differentials to agree with a score cap of 15. 
For example, a 12-8 game is treated as a 15-10 game by multiplying each score by a factor of $1.25$. 
Accordingly, the new score differential of five points is used when calculating ratings.
When the ratings are used to generate score differential predictions (by multiplying the schedule matrix with the rating vector), the predictions are adjusted back to correspond with the original point cap.

This method has distinct qualitative advantages over the USAU method.
It is much simpler to compute, and follows directly from intuition about how game outcomes relate to the strength of the teams involved. 
There are no ad hoc functions or definitions as in the USAU method. 
Furthermore, as we will see below, the least squares system does not require any of the score differential restrictions on eligible games that the USAU method does.

The ratings that this system produces are also practically interpretable: the rating of a team in essence describes by how many points that team is expected to beat an average team, and the difference between two teams tells us the expected point differential if they were to play. 

\section{Empirical comparisons}\label{sec:empiricalcomparisons}

In order to compare the least squares method with the USAU system, we will take advantage of the fact that we can solve \eqref{eq:usau} for the losing score given the winning score, and thus recover a score differential prediction similar to those produced by the least squares.
These score differential predictions can then be evaluated based their performance against the true regular season results.

\subsection{Datasets}

We scraped game data from the USAU website from the 2014 to 2019 seasons for the Club Men's, Club Mixed, and Club Women's divisions. 
The USAU regular season power ratings were also scraped from archived ranking data on the USAU website. 
Archive pages for each club season list all sanctioned regular season and postseason tournaments for each year, and game data is scraped from the corresponding tournament pages. 
Games with missing scores or teams are not recorded, and games with international teams are removed.
Because some information is not available to us after the fact (e.g., unlisted tournaments, roster eligibility concerns), we note that our set of games may differ slightly from the ones used in the official ratings.
We ignore the 2020 and 2021 seasons due to the COVID-19 pandemic and induced irregularity of schedule and teams.

To illustrate the properties of the data and the structure of regular season play, we summarize the games and tournaments for each year and division in Table \ref{tab:data_distribution}. 
The women's division tends to have the smallest number of teams and events, while the mixed division has become the most active division in recent years. 
In all cases, the number of teams, regular season tournaments, and games increased significantly between 2014 and 2019. 
The structure of postseason play has remained the same throughout, so the number of postseason tournaments has stayed constant.

We note here that because the USAU method's data restrictions cause it to rate fewer teams than the least squares method, the sets of games predicted by each are not the same. We simply compare each method based on all predictions that they make, rather than considering only games that both methods consider.

\begin{table}
    \centering
    \resizebox{\columnwidth}{!}{
    \begin{tabular}{clccccc}
    \hline
    \textbf{Year} & \textbf{Division} & \textbf{\# tms.} & \textbf{\# reg. tourn.} & \textbf{\# post. tourn.} & \textbf{\# reg. gms.} & \textbf{\# post. gms.}\\
    \hline \hline
     2014& Men's & 294 & 11 & 31 & 489 & 1013\\ 
    & Mixed & 227 & 14 & 31 & 543 & 1025\\ 
    & Women's & 98 & 8 & 28 & 262 & 391\\ 
    \hline 
     2015& Men's & 263 & 31 & 31 & 1371 & 970\\ 
    & Mixed & 258 & 38 & 31 & 1425 & 1098\\ 
    & Women's & 112 & 23 & 29 & 679 & 376\\ 
    \hline 
     2016& Men's & 280 & 37 & 31 & 1468 & 990\\ 
    & Mixed & 260 & 44 & 31 & 1523 & 1114\\ 
    & Women's & 111 & 20 & 27 & 660 & 374\\ 
    \hline 
     2017& Men's & 294 & 37 & 31 & 1472 & 1048\\ 
    & Mixed & 330 & 44 & 31 & 1713 & 1232\\ 
    & Women's & 124 & 23 & 28 & 763 & 420\\ 
    \hline 
     2018& Men's & 250 & 41 & 31 & 1460 & 993\\ 
    & Mixed & 306 & 51 & 31 & 2097 & 1327\\ 
    & Women's & 116 & 26 & 28 & 710 & 423\\ 
    \hline 
     2019& Men's & 260 & 40 & 31 & 1581 & 1015\\ 
    & Mixed & 339 & 54 & 31 & 2209 & 1323\\ 
    & Women's & 119 & 28 & 29 & 804 & 454\\ 
    \hline
    
    \end{tabular}
    }
    \vspace{0.1in}
    \caption{Summary of regular season structure for each season and division of play, showing the number of teams, regular season tournaments, postseason tournaments, regular season games, and postseason games.}
    \label{tab:data_distribution}
\end{table}

\subsection{Comparative metrics}

As the power ratings are primarily used for determining tournament bids after the regular season concludes, we will compare the accuracy of each rating system according to the predictions that each method makes relative to regular season results.

We will discuss accuracy using multiple metrics. 
The $L^1$ and $L^2$ norms are both appropriate choices to measure the difference between a vector of game predictions and a vector of game outcomes. 
As mentioned above, the two methods predict different numbers of games, so it is more natural to use mean squared error (MSE) and mean average deviation (MAD) to account for differences.

To capture the effectiveness of each method as a \textit{ranking} (ordinal), we will use the ranking violation approach as in \citep{barrow2013ranking} and \citep{coleman2005minimizing}. 
This measures the number of times a lower ranked team, according to a set of ratings, beats a higher ranked team as a fraction of the total games.

\subsection{Results}

In Table \ref{tab:top_25}, we compare the top 25 teams according to each system.
For context, we point out that Seattle Sockeye would go on to win the Men's division national championships, with Chicago Machine in second place and Raleigh Ring of Fire and New York PoNY tied for third place.

\begin{table}
\centering
\resizebox{\columnwidth}{!}{
\begin{tabular}{rlr|rlrr}
 & \multicolumn{1}{c}{\textbf{USAU}}  & \multicolumn{1}{c}{} & & \multicolumn{1}{c}{\textbf{Least Squares}}   & \\ \hline
Rank & Team              & Rating & Rank & Team              & Rating & Diff. \\ \hline \hline
1.    & Seattle Sockeye           & 2304 & 1.  & Seattle Sockeye           & 16.269 & $-$ \\
2.    & New York PoNY              & 2182 & 2.  & D.C. Truck Stop        & 16.075 & $+1$\\
3.    & D.C. Truck Stop        & 2178 & 3.  & San Francisco Revolver          & 16.055 & $+2$\\
4.    & Raleigh Ring of Fire      & 2174 & 4.  & Raleigh Ring of Fire      & 15.283 & $-$\\
5.    & San Francisco Revolver          & 2111 & 5.  & New York PoNY              & 15.076 & $-3$\\
6.    & SoCal Condors     & 2082 & 6.  & Minneapolis Sub Zero          & 14.261 & $+1$\\
7.    & Minneapolis Sub Zero          & 2068 & 7.  & SoCal Condors     & 13.898 & $-1$\\
8.    & Vancouver Furious George    & 2053 & 8.  & Toronto GOAT              & 13.828 & $+1$\\
9.    & Toronto GOAT              & 2014 & 9.  & Chicago Machine   & 13.437 & $+1$\\
10.   & Chicago Machine   & 2013 & 10. & Boston DiG               & 13.338 & $+1$\\
11.   & Boston DiG               & 1974 & 11. & Vancouver Furious George    & 13.154 & $-3$\\
12.   & Denver Johnny Bravo      & 1906 & 12. & Pittsburgh Temper & 12.423 & $+1$\\
13.   & Pittsburgh Temper & 1901 & 13. & Austin Doublewide        & 12.260 & $+1$\\
14.   & Austin Doublewide        & 1885 & 14. & Denver Johnny Bravo      & 12.140 & $-2$\\
15.   & Portland Rhino Slam!       & 1877 & 15. & Atlanta Chain Lightning   & 12.002 & $+1$\\
16.   & Atlanta Chain Lightning   & 1853 & 16. & Cleveland Smokestack    & 10.762 & $+1$\\
17.   & Cleveland Smokestack    & 1791 & 17. & Philadelphia Patrol            & 10.478 & $+1$\\
18.   & Philadelphia Patrol            & 1772 & 18. & Portland Rhino Slam!       & 10.402 & $-3$\\
19.   & Amherst Sprout            & 1749 & 19. & Amherst Sprout            & 10.351 & $-$\\
20.   & Madison Yogosbo           & 1691 & 20. & Madison Yogosbo           & 10.008 & $-$\\
21.   & Seattle Voodoo            & 1674 & 21. & Seattle Voodoo            & 9.894 & $-$\\
22.   & Hampton (VA) Vault             & 1649 & 22. & Indianapolis Brickyard         & 9.822 & $+1$\\
23.   & Indianapolis Brickyard         & 1648 & 23. & Winnipeg General Strike    & 8.981 & $+1$\\
24.   & Winnipeg General Strike    & 1641 & 24. & Hampton (VA) Vault             & 8.700 & $-2$\\
25.   & Dallas Nitro             & 1566 & 25. & Houston Clutch            & 8.282 & $+1$\\ \hline
\end{tabular}
}
\vspace{0.15in}
\caption{Comparison of the top 25 teams by least squares and USAU ratings for the 2019 Club Men's regular season.}
\label{tab:top_25}
\end{table}

As rankings, the two systems are remarkably similar.
All of the top 25 teams are within three places between each method. 
Notably, the least squares method ranks PoNY, Furious George, Johnny Bravo, and Rhino Slam! much lower than the USAU method, while the remaining teams are ranked in a similar order.
In terms of the overall distribution, both methods exhibit a linear decay in rating among the top 25 teams.

The fact that the top five teams by least squares have ratings above 15 is not an error. 
As mentioned above, the least squares rating roughly represents by how many points a team is expected to beat an average team. 
For teams with ratings above 15 points, this implies that those teams would beat both average \textit{and} slightly above-average teams by a full 15 points as well.
This is not unexpected given the talent disparity among teams.

As ratings, the systems do have some notable differences. 
The USAU method has Seattle Sockeye rated over 100 points higher than New York PoNY, which amounts to a full point difference according to \eqref{eq:usau}.
The least squares method, however, has Sockeye favored by only a fifth of a point compared to the next best team.
Further down the list, we see that the USAU method has Toronto GOAT and Chicago Machine separated by only a single rating point, while in least squares they are separated by a just over a third of a point.

\begin{figure}
\centering
    \includegraphics[scale=0.3]{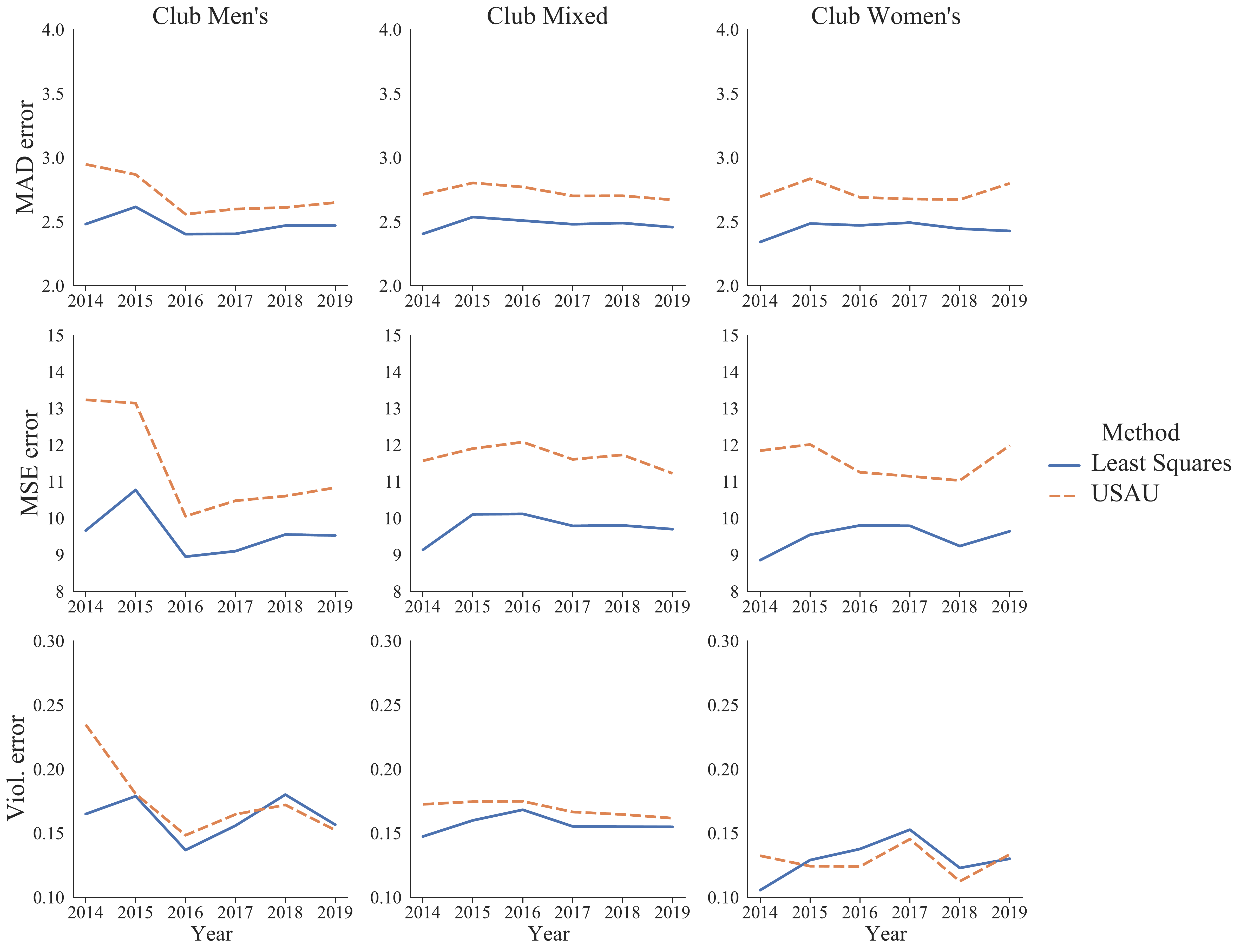}
    \caption{Predictive performance comparison plots for least squares and USAU methods across all years, metrics, and divisions. The least squares method (solid line) demonstrates distinct improvement over the USAU method (dashed line) in MAD and MSE error, and comparable performance with respect to ranking violations.}
    \label{fig:plots}
\end{figure}

We now turn our attention to historical accuracy across all teams. Figure \ref{fig:plots} contains plots of the retrodictive accuracy of each method relative to the metrics described above. 
The least squares method demonstrates a marked improvement over the USAU method relative to mean squared error and mean absolute error, and demonstrates similar performance according to ranking violations. 
In mean absolute error, the least squares method consistently estimates game outcomes to about a quarter point closer than USAU, which is not insignificant across a season. 

The two methods are quite close. 
Aside from the substantial difference for the 2014 Men's division (in which the USAU rankings predict nearly a quarter of all games incorrectly!), both methods are within two percentage points of each other. 
Overall, it is worth reiterating that the least squares method is able to demonstrate improved predictive performance even including the games and teams that the USAU method leaves out. The fact that least squares can consider all data uniformly is a significant advantage.

\section{Conclusions and future work}\label{sec:conclusion}

In this paper, we discussed the sport of ultimate frisbee and properties that make it unique from a sports rating perspective.
We also discussed the current power rating method used by USA Ultimate to rate teams based on regular season play. 
For each system, we computed ratings and score predictions for regular season games, and compared their efficacy relative to the $L^1$ and $L^2$ metrics as well as with respect to ranking violations.
Overall, we saw that the simpler least squares method demonstrated improvement over the current USAU method, and has many desirable qualitative properties.

There is a great deal more work that could be done in this setting. 
Because competitive ultimate until now has not been examined as an application for sports ratings, a comparison of other approaches (such as those mentioned in the introduction) applied to ultimate data would be quite illuminating. 
There are also more divisions available to consider:
USA Ultimate also governs college and youth ultimate, so an investigation into the efficacy of ratings in those settings would be instructive. 
College and youth ultimate possess different structural properties from the club division, which may lead to interesting results.

The unique properties of ultimate also make it a good candidate for the construction of new rating methods.
From the least squares perspective, the Gauss-Markov Theorem tells us that the the ordinary least squares estimator is the best unbiased linear estimator if all observations (in our context, games) have the same variance and the covariance between distinct observations is zero. 
In a sports setting this is of course unreasonable, and so a method to estimate the variance of games may permit the use of a weighted or generalized least squares approach.
Little work has been done on covariance estimation in this setting.
We suspect that a data-driven approach which takes into account additional factors aside from score differential may be able to produce reasonable covariance estimates.
Additional ideas include ratings optimized for least absolute deviation, as well as bounded ratings that take into account score caps directly.
\if0\blind
{
\bigskip
\begin{center}
{\large\bf ACKNOWLEDGEMENTS}
\end{center}

The author would like to thank Dr. Jamie Haddock (Harvey Mudd College) for bringing this method to his attention as well as useful feedback. Will Gilroy, Nathan Hu, Hannah Kaufman, and Austin Froelich also provided helpful discussions.

} \fi
\bigskip
\begin{center}
{\large\bf DECLARATION OF INTEREST}
\end{center}
The author reports that there are no competing interests to declare.
\bigskip
\begin{center}
{\large\bf SUPPLEMENTARY MATERIAL}
\end{center}

\begin{description}

\item[Regular season data:] Regular season game data for 2014-2019 USA Ultimate seasons (zipped folder of .csv files)

\item[Rating code:] Python code used to create and compare the rating methods described in the article. (.py file)

\item[Rating results:] Verbose code output containing game predictions using ratings as well as top 25s for each division and year for both methods described in the article. (zipped folder containing .csv files (top 25's) and .txt (predictions) files).

\end{description}

\bibliographystyle{agsm}
\bibliography{ultimate}

\end{document}